\documentclass{aastex631}

\usepackage{amsmath} 

\def\beq{\begin{equation}\begin{aligned}}
\def\eeq{\end{aligned}\end{equation}} 
\definecolor{LightCyan}{rgb}{0.88,1,1} 
\definecolor{DarkCyan}{rgb}{0.5,1,1} 

\shorttitle{AASTeX v6.3.1 Sample article}
\shortauthors{Brownsberger et al.}
\graphicspath{{./}{figures/}}

\begin{document}

 \title[The Pantheon+ Analysis: Cross Survey Zeropoints and Cosmology]{The Pantheon+ Analysis: Dependence of Cosmological Constraints on Photometric-Zeropoint Uncertainties of Supernova Surveys}

\author[0000-0002-5430-4355]{Sasha Brownsberger}
\affiliation{Department of Physics, Harvard University\\
17 Oxford Street \\
Cambridge, MA 02138, USA}
\email{sashabrownsberger@g.harvard.edu} 

\author[0000-0001-5201-8374]{Dillon Brout}
\affiliation{Department of Physics, Harvard University\\
17 Oxford Street \\
Cambridge, MA 02138, USA}
\affiliation{Center for Astrophysics, Harvard \& Smithsonian\\
60 Garden Street \\
Cambridge, MA 02138, USA}

\author[0000-0002-4934-5849]{Daniel Scolnic}
\affiliation{Department of Physics, Duke University \\
Science Drive \\
Durham, NC 27710, USA}

\author[0000-0003-0347-1724]{Christopher W. Stubbs}
\affiliation{Department of Physics, Harvard University\\
17 Oxford Street \\
Cambridge, MA 02138, USA}
\affiliation{Center for Astrophysics, Harvard \& Smithsonian\\
60 Garden Street \\
Cambridge, MA 02138, USA}

\author[0000-0002-6124-1196]{Adam G. Riess}
\affiliation{Department of Physics and Astronomy, Johns Hopkins University \\
3400 Charles Street \\
 Baltimore, MD 21218, USA }
 \affiliation{Space Telescope Science Institute \\
3700 San Martin Drive \\
Baltimore, MD 21218, USA }



\begin{abstract}

Type Ia supernovae (SNe Ia) measurements of the Hubble constant, H$_0$, the cosmological mass density, $\Omega_M$, and the dark energy equation-of-state parameter, $w$, rely on numerous SNe surveys using distinct photometric systems across three decades of observation.  
Here, we determine the sensitivities of the upcoming SH0ES+Pantheon+ constraints on H$_0$, $\Omega_M$, and $w$ to unknown systematics in the relative photometric zeropoint calibration between the 17 surveys that comprise the Pantheon+ supernovae data set.
Varying the zeropoints of these surveys simultaneously with the cosmological parameters, we determine that the SH0ES+Pantheon+ measurement of H$_0$ is robust against inter-survey photometric miscalibration, but that the measurements of $\Omega_M$ and $w$ are not. 
Specifically, we find that miscalibrated inter-survey systematics could represent a source of uncertainty in the measured value of H$_0$ that is no larger than $0.2$ km s$^{-1}$ Mpc$^{-1}$.
This modest increase in H$_0$ uncertainty could not account for the $7$ km s$^{-1}$ Mpc$^{-1}$  ``Hubble Tension'' between the SH0ES measurement of H$_0$ and the Planck $\Lambda$CDM-based inference of H$_0$.
However, we find that the SH0ES+Pantheon+ best-fit values of $\Omega_M$ and $w$ respectively slip, to first order, by $0.04$ and $-0.17$ per $25$ mmag of inter-survey calibration uncertainty, underscoring the vital role that cross-calibration plays in accurately measuring these parameters. 
Because the Pantheon+ compendium contains many surveys that share low-$z$ Hubble Flow and Cepheid-paired SNe, the SH0ES+Pantheon+ joint constraint of H$_0$ is robust against inter-survey photometric calibration errors, and such errors do not represent an impediment to jointly using SH0ES+Pantheon+ to measure H$_0$ to 1\% accuracy.
\end{abstract}



\section{Introduction} \label{sec:intro} 

Since the first determination by \cite{Perlmutter99, Riess98} that dark energy (DE) is causing the Universe to expand at an accelerating rate, observations of type Ia supernovae (SNe Ia) have been integral in the formulation of the canonical $\Lambda$CDM cosmological model. Predictions based on the $\Lambda$CDM cosmology, which is dominated at our moment in cosmic history by a cosmologically-constant dark energy density ($\Lambda$) and by nonrelativistic, collisionless (`cold') dark matter (CDM), have accurately matched a diversity of cosmological observables, such as the formation of cosmic structure \citep{Eisenstein2005, DES2021}, the rate of the Universe's expansion \citep{Perlmutter99, Riess98, Scolnic18}, and the Cosmic Microwave Background (CMB) \citep{Bennett93, Jones06, Hinshaw13, Planck2020}.

However, local measurements of the contemporary rate of cosmic expansion (the Hubble constant, H$_0$) using SNe Ia \citep{Riess2021} are in approximately 4.5$\sigma$ tension with the $\Lambda$CDM-based inferences of H$_0$ derived from observations of the CMB \citep{Planck2020}.  This inconsistency, dubbed the Hubble Tension (HT), is the source of much debate and has motivated (1) additional methods to produce independent measurements of $H_0$ and (2) investigations into sources of systematic uncertainty that these measurements face. 
 Comprehensive reviews on the matter, such as those of \citet{Verde2019, DiValentino2021}, show a general tension between near-Universe measurements of H$_0$ and early-Universe $\Lambda$CDM-based inferences of H$_0$, with the largest inter-data set tension being that between SH0ES and Planck, owing in part to their relatively small measurement uncertainties.

Here we focus on a critical aspect of the SH0ES H$_0$ measurement: SN Ia cross-survey calibration.
Low-redshift ($z < 0.01$) SNe Ia are critical for measuring H$_0$, as only in such nearby galaxies can the SNe luminosities be tied to collocated Cepheid (or other intermediate standard candle) calibrators.  
However, because SNe Ia occur approximately once every century in a typical galaxy, observations of low-redshift ($z < 0.01$) SNe are volume limited, with the historical frequency of detection being less than one per year.  
Owing to this fundamental infrequency of occurrence, recent analyses \citep{Scolnic18, Riess2021}, in order to obtain a statistically meaningful sample, rely on a heterogeneous catalogue of low-redshift SNe observations spanning three decades and multiple photometric systems usually in the $g$, $r$, $i$ bands.  Though imperfect and idiosyncratic, this painstakingly accumulated catalogue of low-redshift SNe Ia is both critical for SNe Ia-based measurements of H$_0$ and irreplaceable by future surveys. 
In contrast, high-redshift supernovae are much more numerous and considerably fainter, and are best observed by large survey telescopes operating in the $r$, $i$ and $z$ bands with greater sensitivities and wider fields of view. 
Thus, in order to fully utilize all SN Ia distance measurements, contemporary SN Ia analyses combine separate datasets acquired across three decades of observations on many distinct photometric systems.

The calibration of each photometric system used in SN Ia cosmology is performed in each respective data release using heterogeneous methodologies. Surveys such as Pan-STARRS (PS1), Foundation, the Supernova Legacy Survey (SNLS), the Sloan Digital Sky Survey (SDSS), and the Dark Energy Survey (DES) utilize multiple Calspec standards to define their photometry on the AB system and tie nightly photometry to stellar catalogs in each survey’s natural system. 
Some surveys utilize sophisticated forward modeling approaches \citep{Stubbs2006} to infer top of the atmosphere magnitudes of their nightly photometry standards (e.g. UberCal as part of the PS1 analysis: \citealt{Schlafly2012, Scolnic18}, and the Forward Global Calibration Method used in the DES analysis: \citealt{Burke2018}).
Conversely, some low-$z$ samples are tied to the Vega system while others are tied to the AB system using only BD$+17^{\circ}$4708 to tie magnitudes from each filter to the AB system. Furthermore, the low-$z$ surveys' nightly photometric zeropoints are determined differently from those of the larger high redshift surveys. For the low-$z$ surveys this is done either by transforming Landolt \citep{landolt} and Smith standards \citep{smith2002} onto the respective natural system or by transforming the nightly photometry onto the system of Landolt standards. 

Additionally, each respective SNe sample has historically had its own proprietary photometric pipeline. Some use sophisticated forward modeling photometry \citep{holtzmann08,Brout18a,Burke2018}, some use difference imaging \citep{Rest14,Foley17}, and some use difference imaging for SNe that are near their host galaxies while foregoing it when host-galaxy contamination appears minimal \citep{cfa1}. Furthermore, subtle photometric issues have gradually been identified.  
For example, \cite{Kessler2015} and \cite{Brout18a} report the host brightness anomaly, which appears to be worsened by photometric analysis and astrometric uncertainty, and \cite{Scolnic18} identify photometric non-linearities. 
These issues have been addressed in some contemporary surveys, but retroactive correction of some older surveys is impossible since the original data are no longer accessible.

These systematic uncertainties in the flux calibration of the SNe Ia can be distilled into the following aspects: systematic uncertainty in the calibration of standard star observations to the AB  primary standard(s), systematic uncertainties in the set of primary standard(s) themselves used by each survey, systematic uncertainties in the transfer of zeropoints between the local standards near the SN on the night of observations to the primary standards, systematic uncertainties in the measurements of the local standards themselves, and finally systematic uncertainties arising from each photometric pipeline.
The analysis processes applied across the numerous SNe Ia surveys are heterogeneous.  

Stitching many distinct surveys into a unified data set requires cross-calibration, and considerable effort has already been invested to place SNe Ia on the same the flux scale for a self-consistent and homogeneous Hubble diagram.
For example, in the SuperCal calibration method \citep{supercal}, the uniformity of the Pan-STARRS photometric system across a large portion of the sky is used to tie every observed field to background catalogues of reference calibration stars for all 10 surveys used in the original Pantheon Analysis \citep{Scolnic18}. 
Likewise, SuperCal-Fragilistic (Brout et al. in prep.) perform incremental improvements on SuperCal and allow for covariance between surveys including Pan-STARRS, and most notably extend the cross-calibration to include all 17 surveys analyzed in the Pantheon+ analysis (Brout in prep.). 
However, there remain subtleties and assumptions made during the cross-calibration process and it is entirely possible that residual inter-survey photometric zeropoint differences remain. 
As an alternate to SuperCal, \cite{Currie2020} propose the CROSS-CALIB Baysian-based approach to inter-survey calibration, tying tertiary stars across surveys to Pan-STARRS and SDSS objects.  

This exhaustive cross-calibration effort has made possible the compilation of 17 heterogeneous SNe Ia surveys into the single Pantheon+ SNe compendium, which will provide the strongest near-universe constraints of cosmological parameters to-date (Brout in prep.).  
However, it is possible that some undiagnosed inter-survey photometric zeropoint errors remain in Pantheon+.  
By introducing floating, color-independent (gray) flux offsets for each survey, we, in this paper, assess the effect that such residual systematics can have on the on the SH0ES+Pantheon+ joint constraints on H$_0$, on the matter fraction, $\Omega_M$ and on the dark energy equation-of-state parameter, $w$. 
The SH0ES+Pantheon+ joint constraint is obtained by combining an updated version of the SH0ES data set of low-$z$ SNe and calibration Cepheids \citep{Riess2018} with the updated Pantheon+ version of the Pantheon compendium of low-$z$, mid-$z$, and high-$z$ SNe \citep{Scolnic18}. 

Our paper is organized as follows.  In Section \ref{sec:data}, we summarize the Pantheon+ SNe Ia compendium.  In Section \ref{sec:calculations}, we show how we incorporate inter-survey zeropoint offsets into the Pantheon+ SNe Ia data.  In Section \ref{sec:sampling}, we explain the MCMC that we run with these inter-survey zeropoints as free parameters.  
In Section \ref{sec:expandedPosteriors}, we show how the inclusion of conservative but realistic inter-survey zeropoint offsets expands the MCMC posterior distributions and thus the Pantheon+ uncertainties of H$_0$, $\Omega_M$ and $w$.  
In Section \ref{sec:HubbleTension}, we demonstrate that the increase in these posterior uncertainties have clear maximum values that they cannot exceed, no matter how poorly the inter-survey photometric calibration is known.  We explain why these maxima exist in Appendix \ref{app:H0SelfCalibration} and discuss their implications for HT and for SNe Ia measurements of $\Omega_M$ and $w$ in Section \ref{sec:consequences}. 
We conclude our analysis in Section \ref{sec:conclusion}, where we also discuss its immediate implications and possible future applications.  In Section \ref{sec:relStats}, we compare the relative statistical power of adding more SNe to the Pantheon+ data set vs improving our confidence in the existing inter-survey calibrations, and in Section \ref{sec:CSP} we note the vulnerability that a measurement of H$_0$ that relies on a single survey of Hubble Flow (HF) SNe would have to zeropoint inter-survey calibration errors.  

\section{Data \& Methodology} \label{sec:method} 
In this section, we review the upcoming Pantheon+ SNe Ia compendium and explain the methodology that we used to determine the effect that miscalibrated inter-survey zeropoints could have on measurements of cosmological parameters. 
Our analysis is agnostic as to the source of these miscalibrations, assumes that the offsets are gray, and assumes that the offsets between surveys are uncorrelated.  

\begin{figure*}
\centering
\includegraphics[width=0.8\textwidth]{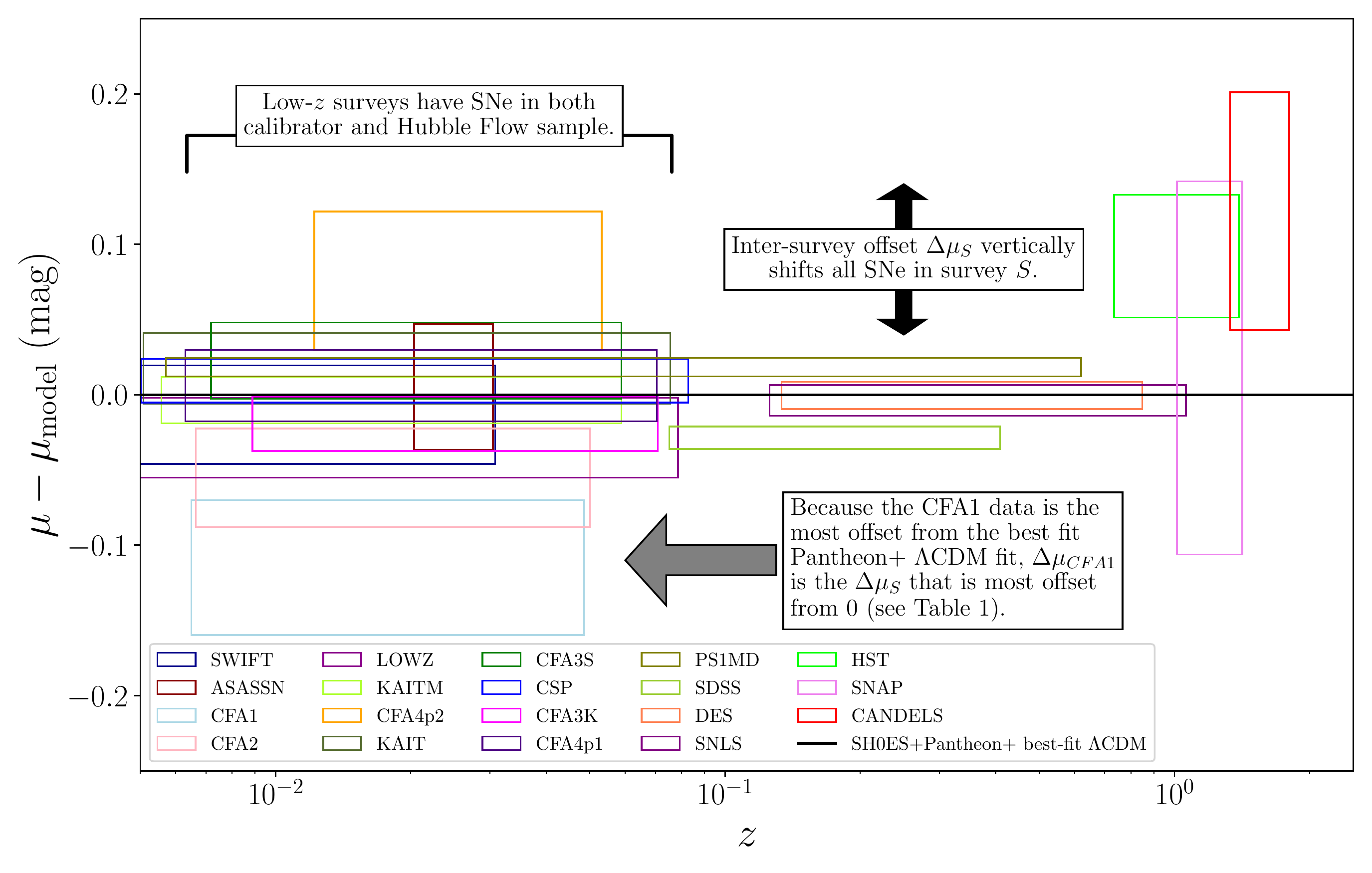}
\caption{\label{fig:PantheonPlusSNe}  The distribution of Pantheon+ SNe Ia in redshift, $z$, and distance modulus residual, $\mu - \mu_{\textrm{model}}$, space, where $\mu_{\textrm{model}}$ is calculated using the best-fit SH0ES+Pantheon+ $\Lambda$CDM cosmology (black line).  
The horizontal extents of the rectangles span the redshift ranges of the surveys, and their vertical extents define the $\pm 1\sigma$ bounds of the the surveys' mean $\mu - \mu_{\textrm{model}}$.  
The Pantheon+ data set (Scolnic et al, in prep.; Brout et al in prep.) is the most expansive collection of spectroscopically-matched SNe Ia ever compiled, containing 1800 SNe gathered from 17 distinct surveys spanning three decades of observations.
To place these SNe on this single plot, the Pantheon+ analysis team undertook an exhaustive cross-calibration effort.  
To assess what effect gray errors in this inter-survey cross-calibration could have on the SH0ES+Pantheon+ cosmological parameter constraints, we introduce, for each survey, a zeropoint offset, $\Delta \mu_S$, as a free parameter that vertically shifts all SNe in that survey.  
Those surveys (CFA1, for example) that are most excursive from the best-fit Pantheon+ $\Lambda$CDM cosmological fit have the largest best-fit $\Delta \mu_S$ values (see Table \ref{tab:MCMCCov}).} 
\end{figure*}

\subsection{Pantheon+ Data Sample}  \label{sec:data} 
We utilize the upcoming Pantheon+ (Scolnic et al. in prep.; Brout et al. in prep.) compendium of publicly available, spectroscopically classified, photometric light curves of SNe Ia that have been cross-calibrated in Brout et al. (in prep.) with redshifts reported by Carr et al (in prep.) and Peterson et al (in prep.). The SNe range in redshift from very low-$z$ ($z<0.01$) used for Cepheid-SN Ia luminosity calibration, to low-$z$ ($0.02<z<0.15$) for filling out the Hubble flow (HF), to high-$z$ ($z>0.15$) usable for measurements of dark energy. The low-$z$ SNe used here are from the Carnegie Supvernova Project (CSP,  \cite{Stritzinger2010}), the Center for Astrophysics SNe Ia data sets (CfA1-4, \cite{cfa1,cfa2,Hicken09b,Hicken09a,Hicken12}), the Lick Observatory Supernova Search (LOSS, \citep{kaitmo, Stahl2019}), the Swift Optical Archive (SWIFT, \cite{swift}), the All Sky Automated Survey for SuperNovae (ASASSN, \cite{asassn}), and the recently released 180 low-$z$ SNe from the Foundation sample \citep{Foley17}. 
At high-z, we include SNe from the first data release of the Pan-STARRS Medium-Deep Field observations (PS1MD, \cite{Rest14,Scolnic18}), SDSS (\cite{Sako11}), SNLS (\cite{Betoule2014}), the DES 3-year sample \citep{Brout18a,Brout18b,Smith_2020}, and the Hubble Space Telescope (GOODS: \citealt{goods1,goods2}, SCP: \citealt{suzuki}, HST: \cite{Riess2018}). 
The distribution of SNe~Ia in redshift and best-fit $\Lambda$CDM residual space can be found in Figure \ref{fig:PantheonPlusSNe}.

The full derivation of distances will be detailed in Brout et al. (in prep.). In short, the light-curve fit parameters of each SN ($m_B$,$c$,$x_1$) are determined using the SALT2 model \citep{Guy2007} as re-trained on the Supercal-Fragilistic system in Brout et al (in prep.). Distances are inferred following the Tripp estimator \citep{Tripp1998}. The distance modulus ($\mu$) to each candidate SN~Ia is obtained by:
\begin{equation}
\label{eq:tripp}
    \mu = m_B + \alpha x_1 - \beta c - M + \delta_{\rm bias}, 
\end{equation}
where $m_B$ is the $B-$band peak-brightness based off of the light-curve amplitude (log$_{10}(x_0)$), where $M$ is the absolute magnitude of a SN~Ia with $x_1=c=0$, and where $\delta_{\rm bias}$ is the bias-correction term that corrects for selection effects. $\alpha_{\rm SALT2}$ and $\beta_{\rm SALT2}$ are the correlation coefficients that standardize the SNe~Ia and are determined following \cite{Marriner11, Popovic2021}.

Typical selection cuts are applied on the observed data sample: we require fitted color uncertainty $< 0.05$, fitted stretch uncertainty $< 1$, fitted light-curve peak date uncertainty $< 2$, light-curve fit probability (from SNANA) $> 0.01$, and we apply Chauvenaut's criterion to the distance modulus residuals, relative to the best-fit cosmological model, at 3.5$\sigma$. 
All SNe data are processed on the same SALT2 light curve fitter.  

\subsection{The Effects of Inter-Survey Zeropoints on SNe Distance Moduli}  \label{sec:calculations} 
We seek to determine the effect that gray inter-survey zeropoint calibration errors could have on the SH0ES+Pantheon+ constraints of cosmological parameters.  
In this Section, we detail how we introduce model agnostic inter-survey zeropoint offsets, representing hypothetical gray miscalibrations, into our analysis.  

For HF SNe Ia $i$ in survey $S$, we adjust its initially determined distance modulus by a new free fit parameter, $\Delta \mu_S$, that applies to all SNe in survey $S$.  
All SNe in survey $S$ are shifted by the same $\Delta \mu_{S}$, of which we have a total of 17, one for each Pantheon+ survey. 
As we derive in Appendix \ref{app:DeltaMuDerivation}, this shifts the measured distance modulus of HF SN $i$ from the initial value of $\mu_{\textrm{meas, init}, i}$ to a shifted value of $\mu_{\textrm{meas, shift}, i}$ according to: 
\beq \label{eq:muShiftDef} 
\mu_{\textrm{meas, shift},i} = \mu_{\textrm{meas, init}, i} - \Delta \mu_L - \Delta \mu_{S} 
\ , 
\eeq 
where $\Delta \mu_L$ is the distance modulus correction that results from our redetermination of the fundamental SNe Ia luminosity, and is the same for every SNe, regardless of survey.  
As we describe in Appendix \ref{eq:fullDeltaMu_app}, we calculate $\Delta \mu_L$ from all $\Delta \mu_S$'s of the Cepheid-paired SNe Ia:  
 \beq \label{eq:fullDeltaMu}
\Delta \mu_L  = - \frac{\sum_{i=1}^{N_{Ceph}} \Delta \mu_{S,i} \sigma_{\mu,i}^{-2} } {\sum_{i=1}^{N_{Ceph}} \sigma_{\mu,i}^{-2} } \ ,
\eeq
where $\sigma_{\mu, i}$ is the reported uncertainty in the distance modulus of SN $i$ and the sum is taken only over Cepheid-paired SNe.  
Including $\Delta \mu_L$ is necessary because adjusting the photometric zeropoint of a Cepheid-paired SN adjusts our inference of that SN's luminosity based on the Cepheid-determined distance to the host galaxy.  

As we detail in Appendix \ref{app:H0SelfCalibration}, because $\Delta \mu_L$ is the negative average inter-survey zeropoint offset of all Cepheid paired SNe, $\Delta \mu_L + \Delta \mu_S$ only shifts the zeropoint of survey $S$ relative to the average shift of surveys with Cepheid-paired SNe.  Applying inter-survey zeropoint offsets (the $\Delta \mu_S$'s) thus cannot shift all  SNe distance moduli collectively, but only relative to each other. 
 

\begin{figure*}
\centering
\includegraphics[width=1.0\textwidth]{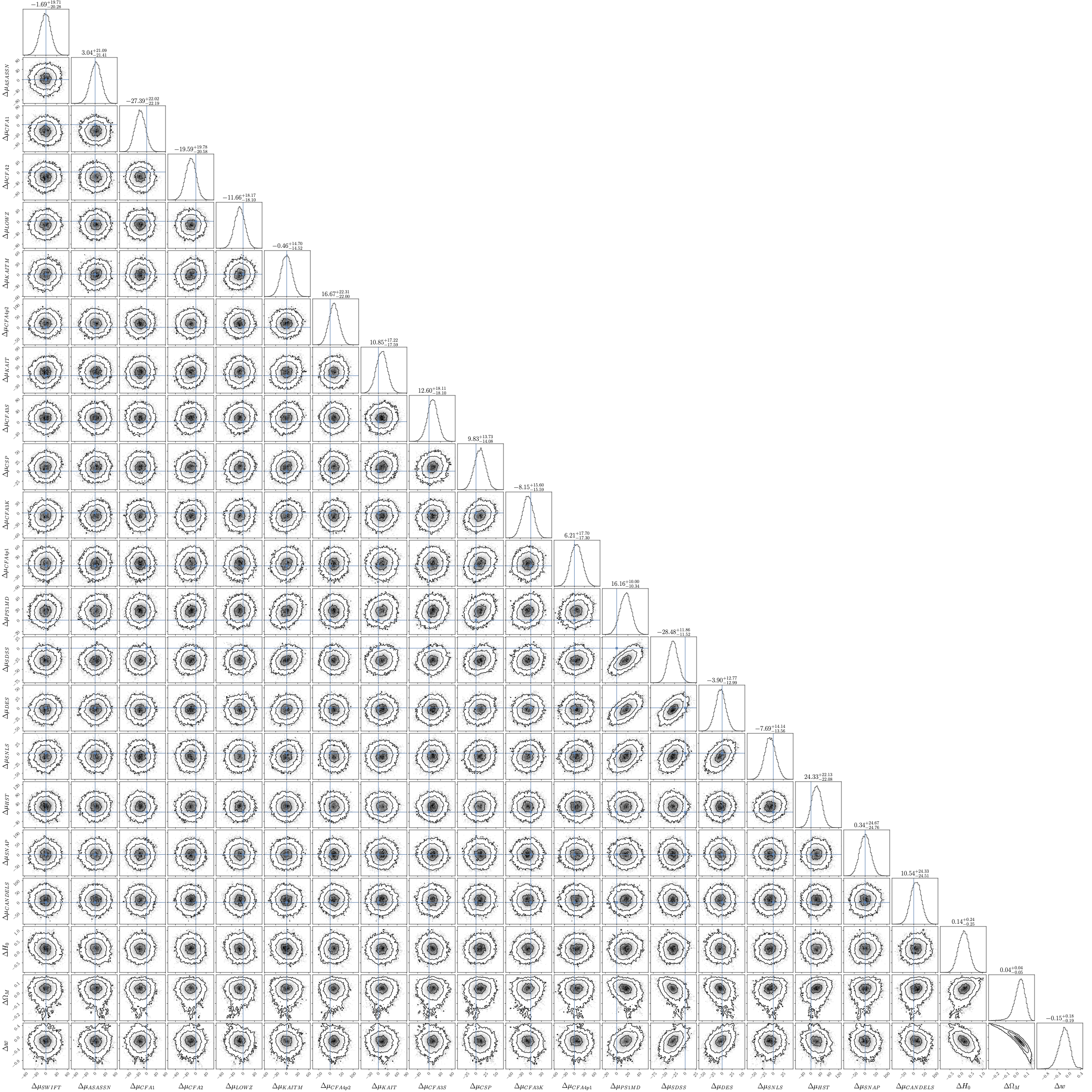}
\caption{\label{fig:fullPosteriorTriangle} The triangle plot showing the posterior distributions of all varied parameters ($\Delta$H$_0$, $\Delta \Omega_M$, $\Delta w$, and one $\Delta \mu_S$ for each survey) with 25 mmag-wide Gaussian priors on the $\Delta \mu_S$ variables.  
The $\Delta \mu_S$'s are given in units of mmag, and $\Delta H_0$ is given in units of km s$^{-1}$ Mpc$^{-1}$ (we omit units in the Figure to provide more space for the labels). 
The $\Delta$'s on the cosmological parameters indicate how much that parameter has shifted from the best-fit value measured when all inter-survey zeropoint offsets are set to 0. 
For the inter-survey zeropoint offsets ($\Delta \mu_S$'s), and blue cross hairs mark $0$, corresponding to total certainty in the Pantheon+ inter-survey photometric calibrations. 
  We list the covariances of H$_0$, $\Omega_M$, and $w$ with all $\Delta \mu_S$'s in Table \ref{tab:MCMCCov}. 
  Circular posteriors indicate weak covariance, while elongated posteriors indicate strong covariance.  
  The covariance between a cosmological parameter and an inter-survey zeropoint offset indicates how much that parameter shifts as the zeropoint of the survey, relative to all other surveys, is varied. 
  The covariance between two inter-survey zeropoint offsets indicates how tightly two zeropoints move together. 
  The tightest covariances are between the largest surveys in the Pantheon+ compendium: DES, SDSS, PS1MD, and SNLS. This tight correlation is unsurprising - separating the zeropoints of two large surveys bifurcates the resulting Hubble diagram, worsening the cosmological fit.}
\end{figure*}

\subsection{Sampling the Space of Cosmological Parameters and Inter-Survey Zeropoint Offsets} \label{sec:sampling} 

To assess what information a data set of SNe Ia measurements provides about cosmological parameters, we compare, for HF SN $i$ the shifted measured distance modulus, $\mu_{\textrm{meas, shift}, i}$, and the theoretically predicted distance modulus, $\mu_{\textrm{model, }i}$.  
As we discuss in Section \ref{sec:calculations}, $\mu_{\textrm{meas, shift}, i}$ shifts when we introduce the inter-survey zeropoint offsets, $\Delta \mu_S$.  
In contrast, $\mu_{\textrm{model, }i}$ is determined entirely by our choice of cosmological parameters and the SN's redshift, $z_i$: 
\beq
 \mu_{\textrm{model, }i}  (z_i; H_0, \Omega_M, w)  
 \  = 25 + 5 \log_{10}  \Big ({\frac{c(1+z_i)}{1 \textrm{Mpc} \times H_0} \int_0^{z_i} \frac{dz'}{H(z', \Omega_M, w)/H_0} }  \Big ) ,  
\eeq
where we assume a flat cosmology and where $c$ is the speed of light.  

We minimize the $\chi^2$ statistic for each of the $N_{HF}$ Hubble flow Pantheon+ supernovae: 
\beq \label{eq:chiSqr} 
\chi^2 = \sum_{i=1}^{N_{HF}} \frac{\Delta \mu_i^2(z_i; H0, \Omega_M, w, L, f_i)}{\sigma_{\mu, i}^2} , 
\eeq
where $\sigma_{\mu,i}^2$ is the variance of $\mu_{\textrm{meas, }i}$ and where we define the predicted vs measured distance modulus residual as 
\beq 
\Delta & \mu_i(z; H0, \Omega_M, w)   = \mu_{\textrm{meas, }i} - \mu_{\textrm{model, i}}(z; H_0, \Omega_M, w) .
\eeq

For supernova $i$ in a survey with zeropoint offset $\Delta \mu_{S}$ with redshift $z_i$ and measured distance modulus $\mu_{\textrm{meas, init, }i}$, the distance modulus residual for a given choice of cosmological parameters and inter-survey offsets is: 
\beq \label{eq:DeltaMu}
\Delta & \mu_i(z_i; \bm{H_0}, \bm{\Omega_M}, \bm{w} ,\bm {\Delta \mu_{S}}) =   \mu_{\textrm{meas, init}, i} - \bm{\Delta \mu_L} - \bm{\Delta \mu_{S}}  - 25  - 5 \log_{10}  \Big ({\frac{c(1+z_i)}{1\textrm{Mpc} \times \bm{H_0}} \int_0^{z_i} \frac{dz'}{H(z', \bm{\Omega_M}, \bm{w}, \bm{H_0})/ \bm{H_0}} }  \Big ), 
\eeq
where the luminosity offset, $\Delta \mu_L$, is calculated from all Cepheid paired SNe and the corresponding values of $\Delta \mu_S$ using Equation \ref{eq:fullDeltaMu}.  For emphasis in Equation \ref{eq:DeltaMu}, we write in boldface the parameters that we vary in the MCMC minimization (note $\Delta \mu_L$ is not directly varied, but is a weighted sum of several varied $\Delta \mu_S$'s).  

To determine the best-fit cosmological parameters (H$_0$, $\Omega_M$ and $w$) and inter-survey zeropoint offsets (the 17 $\Delta \mu_S$'s), we use the \texttt{emcee} MCMC sampling \texttt{Python} library described by \cite{emcee} to minimize $\chi^2$ (Equation \ref{eq:chiSqr}).
We use flat priors to sample the cosmological parameters so that our best-fit posterior values are unbiased.  
However, we use Gaussian priors when sampling $\Delta \mu_S$ to reflect the confidence that the Pantheon+ analysis team has in the accuracy of their inter-survey calibrations (inter-survey zeropoint errors are more likely small than large). 
We center the Gaussian prior for Survey $S$ at $\Delta \mu_S = 0$, and use its width, $\sigma_{P, S}$, to represent our confidence in the Pantheon+ zeropoint calibration of survey $S$.  A prior width of $\sigma_{P, S} <1$ mmag represents effectively complete confidence in the Pantheon+ calibration of survey $S$, and a prior width of $\sigma_{P, S} >100$ mmag represents effectively no knowledge of survey $S$'s zeropoint.  We estimate that a prior width of $\sigma_{P, S} \simeq 25$ mmag on all $\Delta \mu_S$ is a reasonable and conservative representation of the uncertainty of the Pantheon+ cross-calibration (Brout et al. in prep.). 
 
 By running multiple MCMCs with different $\sigma_{P, S}$, we assess the extent to which the conclusions of the standard SH0ES+Pantheon+ analysis (corresponding to setting the $\sigma_{P, S} = 0$) would degrade if the Pantheon+ photometric zeropoints were moderately ($\sigma_{P, S} \leq 25$ mmag), severally ($25$ mmag $< \sigma_{P, S} \leq 50$ mmag), or egregiously ($\sigma_{P, S} > 50$ mmag) miscalibrated .  

\section{Results} \label{sec:results} 

In this Section, we discuss the results of the analysis described in Section \ref{sec:method}.  
\subsection{Updated SH0ES+Pantheon+ Cosmological Parameter Posterior Uncertainties } \label{sec:expandedPosteriors}

\setlength\tabcolsep{5pt}
\begin{table*}
  \centering
  \begin{tabular}{c | c | c | c | c | c | c  | c }
$ \Delta \mu_S $ & $< \Delta \mu_S>$(mmag) & $\frac{d H_0} { d \Delta \mu_S} ( \frac{\textrm{km}}{\textrm{s Mpc 25 mmag }} )$  & $\frac{d \Omega_M } { d \Delta \mu_S} (\frac{1}{\textrm{25 mmag}})$ & $\frac{d w } { d \Delta \mu_S}  (\frac{1}{\textrm{25 mmag}})$ & $N_{\textrm{cal}}$ & $N_{\textrm{HF}}$ & $N_{\textrm{cal}} / N_{\textrm{HF}}$ \\
\hline
$\Delta \mu_{SWIFT}$ & $-1 \pm 20$ & $-0.056 \pm 0.00054$ & $0.000916 \pm 0.00014$ & $-0.00745 \pm 0.00037$ & 14 & 45 & 0.311 \\
$\Delta \mu_{ASASSN}$ & $3 \pm 21$ & $-0.0153 \pm 0.0011$ & $-0.00408 \pm 0.00031$ & $0.0068 \pm 0.00095$ & 4 & 11 & 0.364 \\
$\Delta \mu_{CFA1}$ & $-27 \pm 21$ & $0.00715 \pm 0.00052$ & $-0.00024 \pm 0.00015$ & $-0.00479 \pm 0.00039$ & 2 & 12 & 0.167 \\
$\Delta \mu_{CFA2}$ & $-19 \pm 20$ & $-0.0197 \pm 0.00038$ & $0.00125 \pm 0.00016$ & $-0.0123 \pm 0.00059$ & 4 & 20 & 0.2 \\
$\Delta \mu_{LOWZ}$ & $-11 \pm 18$ & $-0.0273 \pm 0.00038$ & $0.0038 \pm 8.9e-05$ & $-0.0207 \pm 0.00046$ & 8 & 46 & 0.174 \\
$\Delta \mu_{KAITM}$ & $0 \pm 14$ & $-0.0417 \pm 0.00075$ & $0.00211 \pm 0.00017$ & $-0.0186 \pm 0.00061$ & 11 & 93 & 0.118 \\
$\Delta \mu_{CFA4p2}$ & $16 \pm 22$ & $0.00415 \pm 0.00043$ & $0.00224 \pm 0.00014$ & $-0.0121 \pm 0.0004$ & 1 & 11 & 0.0909 \\
$\Delta \mu_{KAIT}$ & $10 \pm 17$ & $0.0028 \pm 0.0012$ & $0.000806 \pm 0.0002$ & $-0.0102 \pm 0.00085$ & 7 & 42 & 0.167 \\
$\Delta \mu_{CFA3S}$ & $12 \pm 18$ & $-0.0162 \pm 0.0014$ & $0.00113 \pm 0.00017$ & $-0.00817 \pm 0.00062$ & 4 & 31 & 0.129 \\
$\Delta \mu_{CSP}$ & $9 \pm 13$ & $0.0372 \pm 0.0015$ & $0.00689 \pm 0.00017$ & $-0.037 \pm 0.0007$ & 10 & 78 & 0.128 \\
$\Delta \mu_{CFA3K}$ & $-8 \pm 15$ & $0.0343 \pm 0.00072$ & $0.0032 \pm 0.0002$ & $-0.0237 \pm 0.00079$ & 4 & 54 & 0.0741 \\
$\Delta \mu_{CFA4p1}$ & $6 \pm 17$ & $0.0294 \pm 0.00055$ & $0.00347 \pm 0.00017$ & $-0.021 \pm 0.00062$ & 1 & 35 & 0.0286 \\
$\Delta \mu_{PS1MD}$ & $16 \pm 10$ & $0.0724 \pm 0.00095$ & $-0.0362 \pm 0.00051$ & $0.153 \pm 0.0016$ & 4 & 442 & 0.00905 \\
$\Delta \mu_{SDSS}$ & $-28 \pm 11$ & $0.0172 \pm 0.001$ & $-0.0479 \pm 0.00039$ & $0.194 \pm 0.00094$ & 0 & 346 & 0.0 \\
$\Delta \mu_{DES}$ & $-3 \pm 13$ & $-0.0176 \pm 0.0015$ & $-0.0295 \pm 0.00028$ & $0.148 \pm 0.00089$ & 0 & 203 & 0.0 \\
$\Delta \mu_{SNLS}$ & $-7 \pm 13$ & $0.0207 \pm 0.00097$ & $0.00981 \pm 0.0002$ & $0.0152 \pm 0.00081$ & 0 & 229 & 0.0 \\
$\Delta \mu_{HST}$ & $24 \pm 22$ & $0.00694 \pm 0.0011$ & $0.0125 \pm 0.00024$ & $-0.0374 \pm 0.00096$ & 0 & 18 & 0.0 \\
$\Delta \mu_{SNAP}$ & $0 \pm 24$ & $0.0126 \pm 0.00039$ & $0.00502 \pm 0.00022$ & $-0.0178 \pm 0.0009$ & 0 & 4 & 0.0 \\
$\Delta \mu_{CANDELS}$ & $10 \pm 24$ & $0.0213 \pm 0.00053$ & $0.00994 \pm 0.00013$ & $-0.0324 \pm 0.00048$ & 0 & 6 & 0.0 \\
  \end{tabular}
  \caption{When the 17 inter-survey offsets, $\Delta \mu_S$, are varied with conservative 25 mmag wide Gaussian priors, we list in this Table their median values (column 2), the slopes of linear fits to the posterior distributions shown in Figure \ref{fig:fullPosteriorTriangle} between each $\Delta \mu_S$ and the three fitted cosmological parameters, H$_0$ (column 3), $\Omega_M$ (column 4), and $w$ (column 5), and the number of calibrator SNe ($N_{\textrm{cal}}$, column 6), the number of Hubble Flow (HF) SNe ($N_{\textrm{HF}}$, column 7), and the  calibrator-to-HF SNe ratio (column 8) in each survey.  
  Each posterior slope (columns 3, 4, and 5) are given as the rate of the parameter change per 25 mmag offset shift.  
We estimate the uncertainty of each fit via statistical bootstrapping. 
  The effect that miscalibration of a survey's zeropoint would have on the best-fit value of a cosmological parameter increases with the magnitude of the corresponding slope.  
  For example, a miscalibration of 25 mmag in the SWIFT zeropoint would shift the best-fit value of H$_0$ by about -0.056 km s$^{-1}$ Mpc$^{-1}$, while a miscalibration of the same size in the SNAP zeropoint would shift H$_0$ by only 0.013 km s$^{-1}$ Mpc$^{-1}$.
  }
  \label{tab:MCMCCov} 
\end{table*}

We first consider the example of a single MCMC run with a uniform, realistic, and conservative prior width of $\sigma_{P,S} = 25$ mmag for all Pantheon+ surveys.  
This choice of $\sigma_{P,S}$ allows the MCMC to accommodate conservative but possible inter-survey photometric miscalibrations while preventing them from expanding unphysically.  
We show the full posterior triangle plot of this MCMC in Figure \ref{fig:fullPosteriorTriangle}.  
In that Figure, the single-parameter histogram for each $\Delta \mu_S$ shows how the relative photometric zeropoint of survey $S$ departs from its Pantheon+ reported value ($\Delta \mu_S = 0$ by definition) when our MCMC allows $\Delta \mu_S$ to float.  

The covariance elements of Figure \ref{fig:fullPosteriorTriangle} indicate how tightly two fit parameters are correlated: circular posteriors indicate little correlation and elongated contours indicate strong correlation. 
We find the tightest correlations between the zeropoint offsets of DES, SDSS, PS1MD, and SNLS.  These are the largest surveys within Pantheon+, and the tight correlations in their offsets are expected: if the zeropoints of these large surveys shift separately, they cannot be easily matched by a single Hubble curve and the overall goodness of fit declines.  

For the data shown in Figure \ref{fig:fullPosteriorTriangle}, we list in Table \ref{tab:MCMCCov} the medians of inter-survey zeropoint offsets and their posterior covariances with H$_0$, $\Omega_M$ and $w$.  
We find that the strength of the dependence of the best-fit cosmological parameters on the inter-survey offsets varies, survey to survey.  
The first-order contribution of a floating survey offset on the shift in the posterior value of a cosmological parameter can be approximated by multiplying the derivative of the parameter with respect to that offset by the median value of that offset. 
For example, we estimate that the the floating zeropoints of SDSS and CFA1 shift the best-fit value of $w$ respectively by: 
\beq
<\Delta w>_{SDSS} & \simeq \frac{d w} {d \Delta \mu_{SDSS}} \Delta \mu_{SDSS} \simeq \frac{0.194 \pm 0.001}{25 \textrm{ mmag}} (-28 \pm 11) \textrm{ mmag}  \simeq -0.22 \pm 0.09 , 
\eeq
 and 
 \beq
 <\Delta w>_{CFA1} & \simeq \frac{d w} {d \Delta \mu_{CFA1}} \Delta \mu_{CFA1}  \simeq \frac{-0.0048 \pm 0.0004}{25 \textrm{ mmag}} (-27 \pm 21) \textrm{ mmag}  \simeq  0.05 \pm 0.04 .
 \eeq 

\begin{figure*}
\centering
\includegraphics[width=0.6\textwidth]{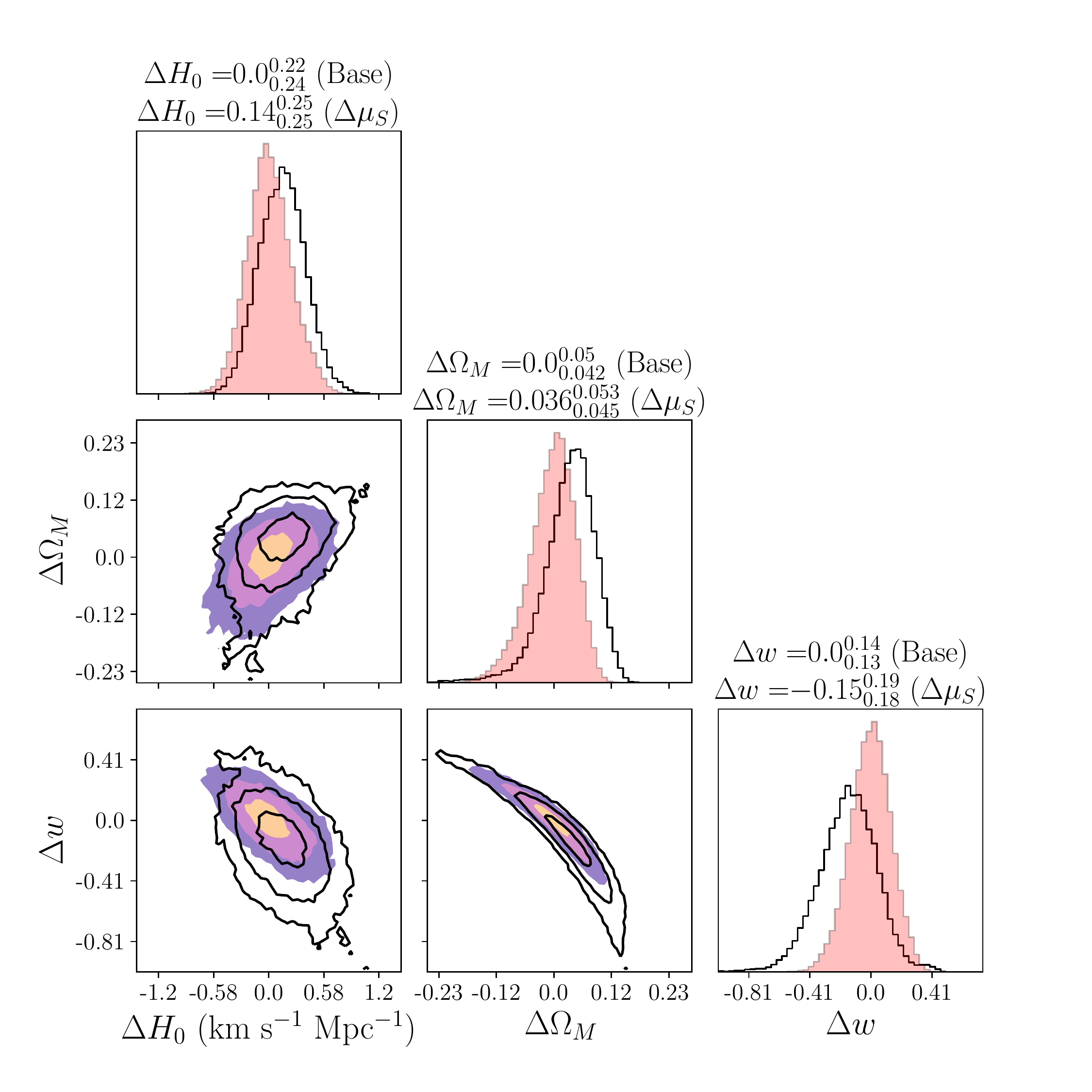} 
\caption{\label{fig:cosmicPosteriorTriangle} The posterior $1\sigma$, $2\sigma$, and $3\sigma$ contours of the $\Delta$H$_0$, $\Delta \Omega_M$, and $\Delta w$ cosmological parameters when the inter-survey zeropoint offsets, $\Delta \mu_S$, are fixed to 0 (filled contours), or allowed to vary with Gaussian priors of widths of $\sigma_{P,S} = 25$ mmag centered at 0 (unfilled contours).  
To maintain the integrity of the SH0ES and Pantheon+ blind, we subtract the median of the posterior distribution of each cosmological parameter when the inter-survey zeropoint offsets are fixed to 0. 
The $\Delta$'s on each parameter indicate that the listed value is measured relative to the base-case ($\Delta \mu=0$) median values.
The inclusion of the $\Delta \mu_S$ as free fit parameters both expand and shift the posterior contours, indicating that they adjust the best-fit cosmological parameter values and increase their reported uncertainties.   
The shifts in H$_0$ are modest, not appreciably relieving HT.  However, the shift in the uncertainties and especially the centroids of $\Omega_M$ and $w$ are substantial. 
Overall, our results suggest that SH0ES+Pantheon+ measurements of H$_0$ are robust against even large inter-survey calibration errors, but that the accuracy of the SH0ES+Pantheon+ reported values of $\Omega_M$ and $w$ are heavily reliant on accurate inter-survey photometric calibration.  
}
\end{figure*}

In Figure \ref{fig:cosmicPosteriorTriangle}, we show the posterior contours of H$_0$, $\Omega_M$, and $w$ when all $\Delta \mu_S$ are fixed to 0 (filled contours), reflecting the default analysis paradigm,
and when all $\Delta \mu_S$ are varied with 25 mmag-wide Gaussian priors centered at 0 (open contours, same as the lower right contours in Figure \ref{fig:fullPosteriorTriangle}).  As expected, the posterior distributions of these cosmological parameters broaden when we introduce additional degrees of freedom in the form of the $\Delta \mu_S$'s.  

The contour shifts in Figure \ref{fig:cosmicPosteriorTriangle} and the nonzero derivatives of the cosmological parameters with respect to the inter-survey offsets in Table \ref{tab:MCMCCov} both indicate that shifting some of the floating $\Delta \mu_S$ values change the posterior Pantheon+ cosmic constraints. 
The shift in the H$_0$ and $\Omega_M$ centroids are modest, less than $1\sigma$ relative to their respective posterior centroids when the $\Delta \mu_S$ offsets are fixed.  
However, the central value of $w$ shifts appreciably, by about $-0.15$ from its null value, which exceeds the null $w$ posterior width of $0.13$.  
Examining columns 2 and 5 of Table \ref{tab:MCMCCov}, this shift does not appear to be the result of a single survey offset.  Rather, the majority of survey offsets impose a negative shift in the posterior value of $w$, and the net effect of these modest shifts is the appreciable shift in the $w$ central value shown in Figure  \ref{fig:cosmicPosteriorTriangle}.  

\subsection{The Dependence of the Cosmological Parameter Constraints on Inter-Survey Zeropoint Offset Uncertainties} \label{sec:HubbleTension} 

To assess what impact systematic errors in the cross-survey calibration could have on the SH0ES+Pantheon+ constraints of cosmological parameters, we run several MCMC analyses with inter-survey zeropoint Gaussian priors, $\sigma_{P, S}$, of varying widths. 
These varying prior widths represent varying degrees of certainty on the accuracy of the Pantheon+ cross-survey calibration.  Priors of $\sigma_{P, S} \leq 10$ mmag represent high confidence in the Pantheon+ inter-survey photometric calibration, priors of $ 10 < \sigma_{P, S} \leq 25$ represent reasonable confidence, priors of $ 25 < \sigma_{P, S} \leq 50$ mmag represent extremely conservative but not outlandish calibration uncertainties, and priors of $\sigma_{P, S} > 50$ mmag represent unrealistically large uncertainties in the Pantheon+ cross-calibration.  

\begin{figure*}
\centering
\includegraphics[width=0.9\textwidth]{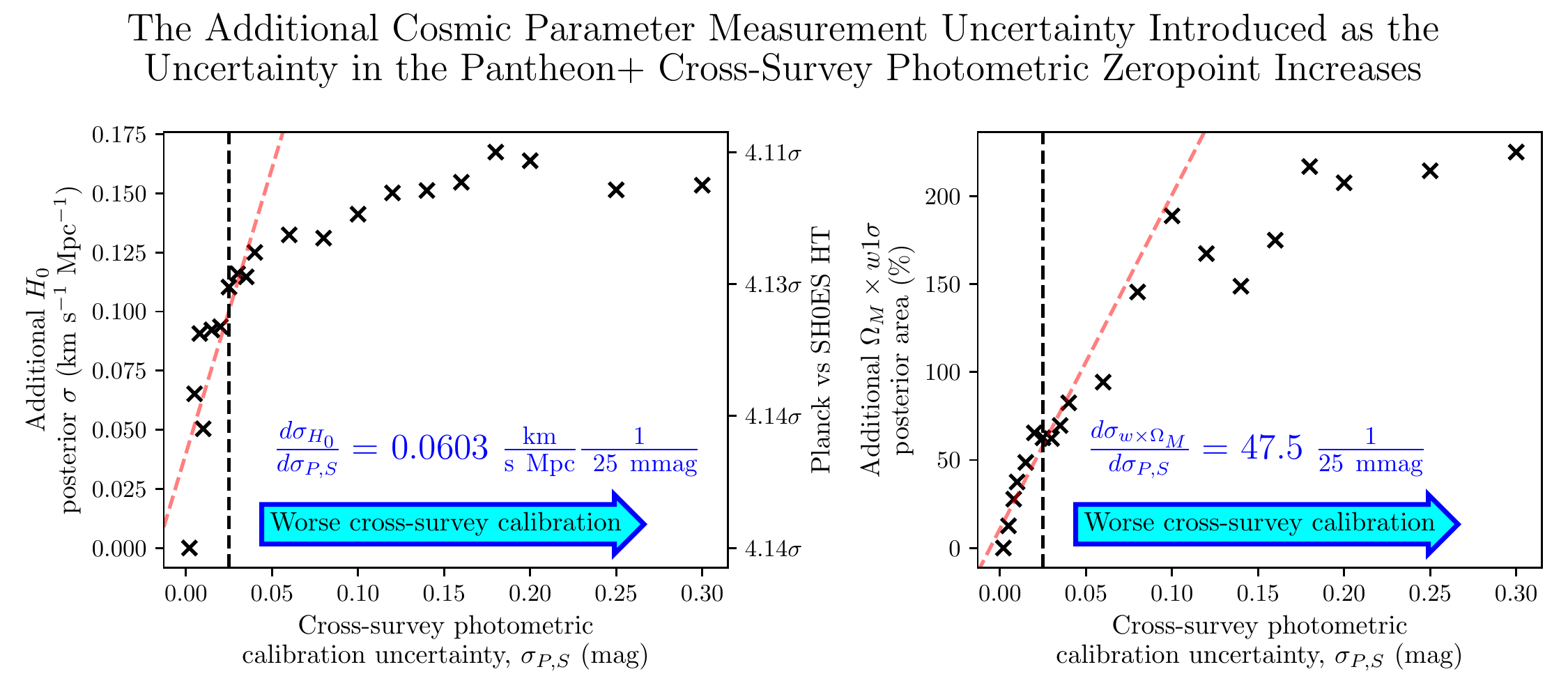}
\caption{\label{fig:PostSigOnGaussPrior} 
The dependencies of the SH0ES+Pantheon+ H$_0$ uncertainty (left) and of the SH0ES+Pantheon+ joint $\Omega_M$ and $w$ uncertainty (right) on the widths of the inter-survey calibration offsets' MCMC Gaussian priors, $\sigma_{P,S}$.  
Larger $\sigma_{P,S}$ correspond to less confidence in the reported Pantheon+ relative photometric zeropoints. 
The dashed black line indicates a  conservative estimate of the worst reasonable Pantheon+ inter-survey calibration error of $25$ mmag. 
A $\Delta \mu_S$ Gaussian prior width of 0 mags is equivalent to running the MCMC with no inter-survey offsets ($\sigma_{P, S}$ fixed to 0 for each survey $S$).  
The posterior widths increase approximately linearly as the $\sigma_{P,S}$ expands from 0 to 40 mmag, increase more slowly for priors between $40 < \sigma_{P,S} \leq 200$ mmag, and largely cease to change at all for $\sigma_{P,S} > 200$ mmag.  
As we explain in Appendix \ref{app:H0SelfCalibration}, this plateauing effect occurs because inter-survey zeropoint offsets can source only relative differences in distance moduli between surveys, and these relative differences can only grow so large before they became inconsistent with a single Hubble curve.
We estimate the realistic dependence of the SH0ES+Pantheon+ measurements on the accuracy of the inter-survey photometric calibrations by fitting a line to the $\sigma_{P,S} \leq 40$ mmag data points, shown as a dashed red line.  
The slopes, which we write in the plots, estimate the rate at which the SH0ES+Pantheon+ parameter measurements slip per $25$ mmag of inter-survey calibration error.  
The right vertical axis of the H$_0$ plot (left) shows the HT between the SH0ES measurement of H$_0$ \citep{Riess2021} and the Planck inference of H$_0$ \citep{Planck2020} when the additional statistical uncertainty introduced by the $\Delta \mu_S$ offsets (left vertical axis) is added in quadrature with the statistical uncertainties reported by SH0ES and Planck.  The change in HT is marginal, reducing the number of statistical $\sigma$ separating SH0ES and Planck measurements of H$_0$ from $4.14\sigma$ to $4.11\sigma$. 
gray inter-survey calibration errors cannot be the source of HT, as they reduce the statistical discrepancy between SH0ES and Planck only negligibly.  
}
\end{figure*}

\begin{figure*}
\centering
\includegraphics[width=1.0\textwidth]{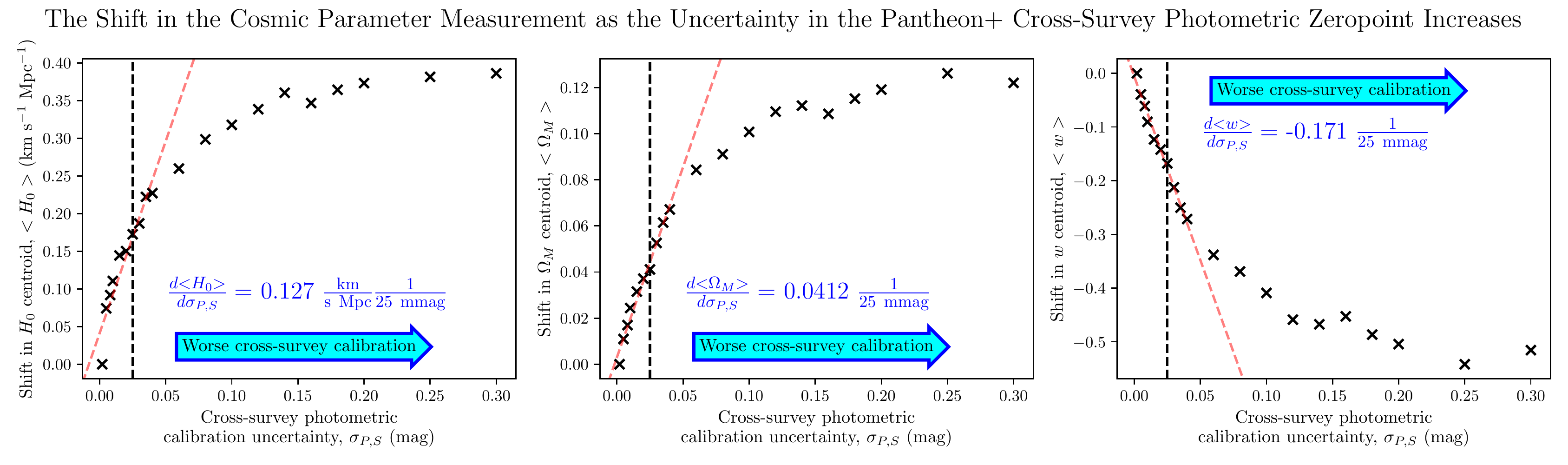} 
\caption{\label{fig:PostCentrOnGaussPrior} 
Same as Figure \ref{fig:PostCentrOnGaussPrior}, except the change in the SH0ES+Pantheon+ $H_0$, $\Omega_M$, and $w$ centroids, indicated with $<\cdot>$, as we increase $\sigma_{P,S}$.  
The slopes of the $\sigma_{P,S} \leq 40$ fitted lines estimate the realistic dependencies of the SH0ES+Pantheon+ best-fit cosmological parameter values on the accuracy of the inter-survey photometric zeropoint calibrations.  
The slippage in $<H_0>$ of 0.13 km s$^{-1}$ Mpc$^{-1}$ $(25$ mmag$)^{-1}$ is negligible, underscoring the robustness of the SH0ES+Pantheon+ measurement of H$_0$ to cross-survey calibration errors.  
However, the $<\Omega_M>$ and $<w>$ slopes of 0.04 $(25$ mmag$)^{-1}$ and -0.17 $(25$ mmag$)^{-1}$ respectively are quite substantial, representing meaningful changes in the class of cosmology measured by SH0ES+Pantheon+.  
We do note that contemporary measurements of $\Omega_M$ and $w$ typically leverage orthogonal cosmic probes, such as SNe Ia and the CMB (see, for example, \citet{Scolnic18}). 
The slip in the joint CMB+SH0ES+Pantheon+ measurements of $\Omega_M$ and $w$ as $\sigma_{P,S}$ increases would be less than the slopes we measure, as the CMB would provide additional stabilization.  
Our results do not indicate that the SH0ES+Pantheon+ measurements of cosmological parameters are inaccurate, but rather that their reliability is highly dependent on the reliability of the Pantheon+ inter-survey calibrations, underscoring the importance of the painstaking cross-calibration work of the Pantheon+ analysis team.  }
\end{figure*}


The cosmological parameters' (H$_0$, $\Omega_M$, $w$) posterior uncertainties expand when the inter-survey zeropoint offsets ($\Delta \mu_S$) are allowed to vary with Gaussian priors of widths specified by $\sigma_{P,S}$.
In Figures \ref{fig:PostSigOnGaussPrior} and \ref{fig:PostCentrOnGaussPrior}, we respectively show the dependence of the posterior widths and centroids of the H$_0$, $\Omega_M$, and $w$ posterior widths on $\sigma_{P,S}$. 
For $\sigma_{P, s} < 40$ mmag, the cosmological parameter posteriors expand and shift rapidly.  As the priors broaden from 40 mmag to about 200 mmag, allowing for progressively larger zeropoint calibration errors, the parameter posterior uncertainties and centroids continue to shift, but at a slower pace. 
Past this 200 mmag threshold the expansions of the MCMC posteriors plateau, remaining approximately constant even as the widths of the priors on $\Delta \mu_S$ increase.  
As we further detail in Appendix \ref{app:H0SelfCalibration}, the combination of many distinct surveys of overlapping redshift ranges spanning both Cepheid calibrators and HF SNe prevents the individual survey zeropoints from drifting apart while remaining consistent with a single set of cosmological parameters.  
This effective `self-calibration' is the source of the plateauing effect in Figure \ref{fig:PostSigOnGaussPrior}, and places a limit on how large an impact inter-survey calibration errors could have on the SH0ES+Pantheon+ measurement of cosmological parameters, even if the true inter-survey zeropoints are effectively unknown ($\sigma_{P,S}>100$ mmag in our analysis).  

Generically, additional surveys introduce additional, independent measurements of H$_0$ and therefore more resilience against inter-survey calibration errors.  
However, because the calibration methods of several surveys are correlated, not every additional survey introduces a totally independent H$_0$ probe.  
There are, therefore, diminishing returns as additional surveys are introduced unless those new surveys are wholly independent from the surveys that came before.  

Even the most conservative estimates of the Pantheon+ inter-survey photometric calibration uncertainty is less than $50$ mmag, or within the $\sigma_{P,S}$ range at which the cosmological parameters posteriors depend approximately linearly on $\sigma_{P,S}$.  
To estimate the realistic dependence on the SH0ES+Pantheon+ measurements of cosmological parameters on the accuracy of the compendium's inter-survey calibrations, we fit lines to the dependencies of the H$_0$, $\Omega_M$, and $w$ posterior widths and centroids on the inter-survey calibration priors, $\sigma_{P,S}$ (red dashed lines in Figures \ref{fig:PostSigOnGaussPrior} and \ref{fig:PostCentrOnGaussPrior}).  
The slopes of these lines, shown in Figures \ref{fig:PostSigOnGaussPrior} and \ref{fig:PostCentrOnGaussPrior}, estimate the sensitivities of the SH0ES+Pantheon+ cosmological parameters constraints to errors in inter-survey zeropoint calibrations.  

\subsection{Implications for HT and SNe Ia Measurements of Cosmological Parameters} \label{sec:consequences}

The linear change of the SH0ES+Pantheon+ H$_0$ uncertainty and centroid on realistic ($\sigma_{P,S} < 40$ mmag) inter-survey zeropoint calibration errors are respectively 0.06 km s$^{-1}$ Mpc$^{-1}$ (25 mmag)$^{-1}$ and 0.13 km s$^{-1}$ Mpc$^{-1}$ (25 mmag)$^{-1}$. 
Even the inclusion of effectively free-floating ($\sigma_{P,S} > 200$) inter-survey offsets increases the SH0ES+Pantheon+ posterior uncertainty of H$_0$ by 0.15 km s$^{-1}$ Mpc$^{-1}$ and shifts the H$_0$ posterior centroid by 0.37 km s$^{-1}$ Mpc$^{-1}$. 
This change in H$_0$ posterior width and centroid are both small, indicating that the SH0ES+Pantheon+ measurement of H$_0$ is quite robust against even egregious inter-survey miscalibrations.  

Using the most updated SH0ES measurement \citep{Riess2021} of H$_0 = 73.2 \pm 1.3$ km s$^{-1}$ Mpc$^{-1}$ as a proxy (the SH0ES best-fit value remains blinded, pending the publication of Brout et al. in prep.), this increase in the H$_0$ uncertainty negligibly enlarges the H$_0$ uncertainty from $1.30$ km s$^{-1}$ Mpc$^{-1}$ to $\sqrt{1.30^2 + 0.16 ^ 2} \simeq 1.31$ km s$^{-1}$ Mpc$^{-1}$. 
This would correspond to a reduction in HT between the Cepheid+SNe Ia distance ladder measurement and the Planck $\Lambda$CDM-based prediction of H$_0$ ($67.27 \pm 0.6$ km s$^{-1}$ Mpc$^{-1}$) from $4.14\sigma$ to $4.11\sigma$, and would not relieve HT.  
In principle, this calculation could also include the centroid shift in H$_0$ (see Figure \ref{fig:PostCentrOnGaussPrior}) which would (because the centroid shift in H$_0$ is positive) actually marginally increase HT.  We opt to take the more conservative approach, only including the increase in the H$_0$ posterior uncertainty.  
Inter-survey photometric calibration errors in Pantheon+ cannot appreciably relieve HT.  

Because the SH0ES+Pantheon+ joint data set contains many low-$z$ surveys that include both Cepheid-paired and HF SNe Ia, the data set's effective `self-calibration' makes the SH0ES+Pantheon+ constraint on H$_0$ quite robust against inter-survey systematic errors.  
As we detail in Appendix \ref{app:H0SelfCalibration}, calibration errors in surveys with Cepheid-paired SNe do not strongly shift the best-fit value of H$_0$ because the shift is compensated for by the correction to the inferred SNe Ia absolute luminosity.  Further, photometric offsets in surveys with only HF SNe can grow only so large before the surveys' distance modulus residuals split at overlapping redshifts in the Hubble diagram, degrading the resulting goodness of fit.  

However, because Pantheon+ has fewer high-redshift ($z>0.4$) SNe Ia and SNe Ia surveys, where changes in $\Omega_M$ and $w$ are most significant, the self-calibration that makes the SH0ES+Pantheon+ measurements of H$_0$ so robust against inter-survey photometric miscalibration does not meaningfully stabilize SH0ES+Pantheon+ measurements of $\Omega_M$ nor of $w$. 
SNe Ia measurements of $\Omega_M$ and $w$ are somewhat degenerate, and we therefore report in Figure \ref{fig:PostCentrOnGaussPrior} the area of the 1$\sigma$ contour in the $\Omega_M \times w$ posterior space as the joint uncertainty in these two parameters.  
As we show in the right panel of Figure \ref{fig:PostSigOnGaussPrior}, effectively free-floating inter-survey offsets ($\sigma_{P, S} >200$) increases the $\Omega_M \times w$ joint posterior uncertainty by 150\%.  In the center and right panels of Figure \ref{fig:PostCentrOnGaussPrior}, the centroids of $\Omega_M$ and $w$ respectively shift by as much as 0.12 and -0.5 if the inter-survey zeropoint calibration is effectively unconstrained.  
Focusing on the more realistic ($\sigma_{P, S} \leq 25$) inter-survey calibration errors, an increase in inter-survey calibration uncertainty of 25 mmag produces slips of, respectively, 0.04 and -0.17 in the SH0ES+Pantheon+ best-fit values of $\Omega_M$ and $w$ (center and right panels of Figure \ref{fig:PostCentrOnGaussPrior}) and of 50\% in their joint posterior uncertainty (right panel of Figure \ref{fig:PostSigOnGaussPrior}).  
These best-fit shift and posterior uncertainty increases are large, substantially larger than the corresponding changes in H$_0$.  They would, if realized,  represent dramatic changes in the class of cosmology measured by SH0ES+Pantheon+. 
Reliable SNe Ia compendium-based measurements of $\Omega_M$ and $w$ therefore rely heavily on accurate inter-survey photometric calibration. 

Notably, gray offsets are sufficient for H$_0$ measurements, which rely on the comparison of ultra low-$z$ ($0.01 < z$) to low-$z$ ($0.1 < z$) SNe at roughly the same colors.  However, our analysis does not encompass possible chromatic inter-survey systematics that could subtly effect SH0ES+Pantheon+ measurements of $\Omega_M$ and $w$.



\section{Discussion and Conclusions} \label{sec:conclusion}

\subsection{Statistical Power of More Surveys vs Improved Inter-Survey Calibrations} \label{sec:relStats}
Ongoing efforts to leverage SNe Ia to gain ever more accurate measurements of the Universe's recent expansion could pursue one of two broad tracks: improved calibration of existing data sets or acquisition of new observations, introducing more data at the cost of additional inter-survey systematics.  
With the hope of providing insight into how limited scientific resources could most efficiently be used, we use artificial SNe data to study how measurements of cosmological parameters improve with additional data vs with improved knowledge of inter-survey systematics.  

We produce artificial low-$z$, mid-$z$, and high-$z$ surveys by randomly drawing 300 SNe logarithmically distributed over redshift ranges of $0<z<0.1$, $0.1<z<0.4$, and $0.4<z<1$, respectively, with associated $\mu$ values determined from the best-fit SH0ES+Pantheon+ cosmology, subject to the median $\mu$ uncertainty of the real Pantheon+ SNe Ia in this redshift range.  
By repeating the analysis discussed Section \ref{sec:sampling} on the Pantheon+ data with the inclusion of each of these artificial surveys, we determine how the SH0ES+Pantheon+ constraints of H$_0$, $\Omega_M$, and $w$ depend on the redshift distribution of SNe added to the Pantheon+ compendium and on the uncertainties of the Pantheon+ inter-survey photometric zeropoints. We list these results in Table \ref{tab:ToyData}. 


\setlength\tabcolsep{0pt}
\begin{table*}
  \centering
  \begin{tabular}{c | c | c | c | c    }
\hline
\hline
Relative H$_0$ uncertainty when: & \cellcolor{LightCyan} $\ $Just Pantheon+ $\ $ & $\ $+ Low-$z$ Survey $\ $ & $\ $ + Mid-$z$ Survey $\ $ & $\ $ + High-$z$ Survey $\ $ \\
\hline
$\sigma_{P,S}  =  5$ mmag & \cellcolor{LightCyan} 95.5\% & 90\% & 97.8\% & 92.9\% \\
\rowcolor{LightCyan} $\sigma_{P,S}  =  10$ mmag & \cellcolor{DarkCyan} 100\% & 94.6\% & 100\% & 94.1\% \\
$\sigma_{P,S}  =  25$ mmag & \cellcolor{LightCyan} 107\% & 101\% & 107\% & 361\% \\
$\sigma_{P,S}  =  50$ mmag & \cellcolor{LightCyan} 112\% & 105\% & 105\% & 103\% \\
\hline
Relative $\Omega_M \times w$ area when: & \cellcolor{LightCyan} Just Pantheon+ & + Low-$z$ Survey & + Mid-$z$ Survey & + High-$z$ Survey \\
\hline
$\sigma_{P,S}  =  5$ mmag & \cellcolor{LightCyan} 87.4\% & 91.7\% & 85\% & 53.8\% \\
\rowcolor{LightCyan} $\sigma_{P,S}  =  10$ mmag & \cellcolor{DarkCyan} 100\% & 106\% & 99.3\% & 63.3\% \\
$\sigma_{P,S}  =  25$ mmag & \cellcolor{LightCyan} 148\% & 133\% & 137\% & 501\% \\
$\sigma_{P,S}  =  50$ mmag & \cellcolor{LightCyan} 165\% & 167\% & 156\% & 114\% \\
\hline
$\ $ Relative $w$ uncertainty reduction when: $\ $ & \cellcolor{LightCyan} Just Pantheon+ & + Low-$z$ Survey & + Mid-$z$ Survey & + High-$z$ Survey \\
\hline
$\sigma_{P,S}  =  5$ mmag & \cellcolor{LightCyan} 91.2\% & 93.7\% & 83.9\% & 64.4\% \\
\rowcolor{LightCyan} $\sigma_{P,S}  =  10$ mmag & \cellcolor{DarkCyan}  100\% & 102\% & 97.8\% & 75.6\% \\
$\sigma_{P,S}  =  25$ mmag & \cellcolor{LightCyan} 131\% & 110\% & 124\% & 1640\% \\
$\sigma_{P,S}  =  50$ mmag & \cellcolor{LightCyan} 134\% & 128\% & 136\% & 95.7\% \\
\end{tabular} 
\caption{The changes in the SH0ES+Pantheon+ cosmological parameter posterior constraints for various inter-survey zeropoint offset prior widths ($\sigma_{P,S}$, different rows) and with the inclusion of low-$z$ ($0.01 < z < 0.1$), mid-$z$ ($0.1 < z < 0.4$), and high-$z$ ($0.4 < z < 1.0$) artificial surveys (different columns). 
Each survey contains 300 `new' SNe, randomly distributed logarithmically in the corresponding redshift range. 
We express these constraints as percentages of the `gold-standard' scenario, with an impressive $\Delta \mu_S$ prior of 10 mmag and with the Pantheon+ compendium as it presently exists (the darkly shaded cyan cells). 
Smaller percentage values indicate tighter constraints.  
As expected, the posterior uncertainties decline as we add surveys and increase as the inter-survey offset priors grow.  H$_0$ is better constrained by the addition of low-$z$ SNe, while $\Omega_M$ and $w$ are better constrained with the addition of high-$z$ SNe.   
The constraining power of a well-chosen survey of 300 SNe is comparable to an improvement in the inter-survey zeropoint cross-calibration confidence from $25$ mmag to $10$ mmag.  
For example, adding a low-$z$ survey of 300 SNe would improve the SH0ES+Pantheon+ constraint on H$_0$ by about 6\% (from 100\% to 94.6\% in this Table) if the inter-survey zeropoints are calibrated to an accuracy of 10 mmag.  
Similarly, improving the inter-survey zeropoint calibration uncertainty from 25 mmag to 10 mmag would improve the constraining power of the existing SH0ES+Pantheon+ data by about 7\% (from 107\% to 100\% in this Table).  
Accurate inter-survey cross-calibration is essential if surveys of SNe Ia are to be fully leveraged.} \label{tab:ToyData} 
\end{table*} 

Examining Table \ref{tab:ToyData}, we find, as expected, that the uncertainties in the best-fit cosmological parameter values decrease both as we add SNe of an appropriate redshift range, and as we decrease the uncertainty (i.e. Gaussian-prior width) of the inter-survey zeropoint offsets.  
The posterior uncertainty in H$_0$ would be improved by about $6-7$\% either with the inclusion of a low-$z$ survey of 300 SNe or with an improvement in the Pantheon+ inter-survey zeropoint calibration from 25 mmag to 10 mmag.  

These results underscore the importance of good inter-survey cross-calibration.  With a compendium of nearly 2000 SNe Ia, a modest improvement in inter-survey calibration (from 25 mmag to 10 mmag uncertainties, for example) provides as much additional cosmological insight as an entirely new, independent survey of 300 additional SNe Ia. 

\subsection{Assessing the Impact of Single Hubble Flow Survey Analysis} \label{sec:CSP}
The Carnegie Chicago Hubble Program (CCHP) measures H$_0$ using only HF SNe Ia observed as part of the Carnegie Supernova Project (CSP) low to intermediate redshift ($z<0.1$) survey \citep{Krisciunas2017, Phillips2019} anchored to a very low-$z$ SNe observed in a variety of surveys tied to tip of the red giant branch (TRGB) measured galactic distances.  
\begin{figure*}
\centering
\includegraphics[width=0.8\textwidth]{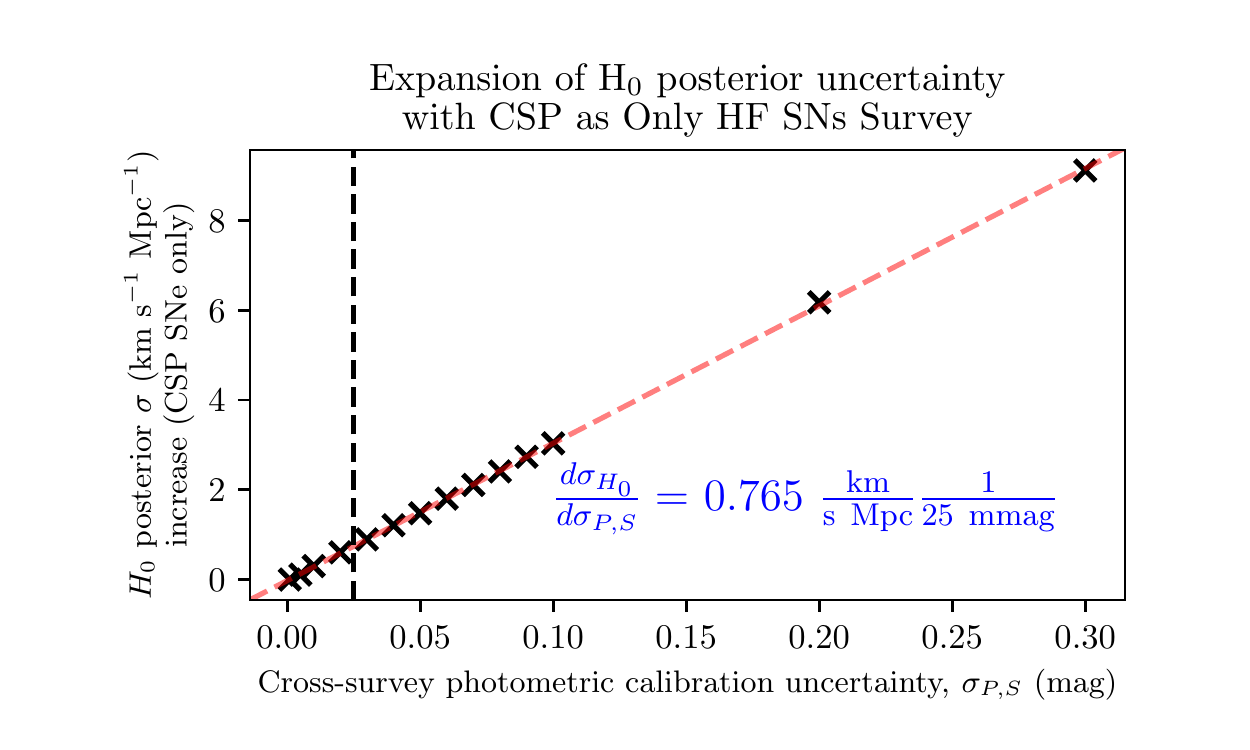}
\caption{\label{fig:CSPH0Posteriors} The dependence of the Hubble Constant, H$_0$, posterior uncertainty on the width of the $\Delta \mu_S$ Gaussian priors when we use only HF SNe Ia detected by the Carnegie Supernova Project (CSP).  
In contrast to the plateauing effect in Figure \ref{fig:PostSigOnGaussPrior}, the posterior uncertainty on the measured value of H$_0$ increases without bound as we increase the width of the $\Delta \mu_S$ priors.  
Because we use HF SNe from only CSP in this analysis, a single cross-calibration error between this HF survey and the other surveys with Cepheid-paired SNe shifts all HF $\mu$ values, and thus the inferred value of H$_0$, without a statistical penalty, leading to a posterior H$_0$ uncertainty that increases effectively linearly with $\sigma_{P, S}$ with a slope of $0.77$ km s$^{-1}$ Mpc$^{-1}$ per every 25 mmag of Pantheon+ calibration uncertainty. 
Focusing on the conservative $\sigma_{P,S} = 25$ mmag estimation of the Pantheon+ cross-calibration uncertainty (dashed black line), the H$_0$ posterior uncertainty increases by $0.77$ km s$^{-1}$ Mpc$^{-1}$, or a full 1\% of the best-fit H$_0$ value, potentially challenging the goal of measuring H$_0$ to better than 1\% accuracy. 
This underscores the importance of accurate inter-survey photometric calibration, as the additional H$_0$ uncertainty depends linearly on the accuracy of the inter-survey zeropoint calibration. }
\end{figure*}

CCHP was initiated in part to sidestep systematic errors in distance ladder measurements of cosmological parameters that could arise from cross-calibrating heterogeneous observations of HF SNe Ia.  
However, because observations of TRGB-paired SNe Ia are volume limited (direct observations of TRGB is only possible in nearby host galaxies with an observed SNe Ia, of which there are a volume-limited number), CCHP is still forced to use numerous optical systems to establish the normalized standard candle luminosity of SNe Ia.  
Cross-calibration errors, therefore, are still present in CCHP. 

In this work, we have thus far demonstrated that the inclusion of multiple surveys with HF SNe makes the resulting measurement of H$_0$ inherently robust against large inter-survey calibration errors.  
In Figure \ref{fig:PostSigOnGaussPrior}, the increase in the posterior H$_0$ uncertainty has a clear maximum value, no matter how large the inter-survey zeropoints might be. 

To approximate the analysis of CCHP, we repeat the analysis of Section \ref{sec:HubbleTension} using all Cepheid-paired SNe Ia but only CSP HF SNe Ia.  Because the CSP SNe Ia in the Pantheon+ compendium are all below redshift 0.08, they do not provide strong constraints on $\Omega_M$ nor on $w$. 
We therefore fix those cosmological parameters to $\Omega_M = 0.3$ and $w=-1$, varying only $H_0$ and the relevant inter-survey zeropoint offsets, $\Delta \mu_S$, in the MCMC.  

We present the dependence of the H$_0$ posterior uncertainty on the Gaussian prior width, $\sigma_{P,S}$, of all $\Delta \mu_S$ in Figure \ref{fig:CSPH0Posteriors}. Notably, and in contrast to the results of Figure \ref{fig:PostSigOnGaussPrior}, Figure \ref{fig:CSPH0Posteriors} shows no plateauing in the uncertainty on $H_0$.  
Thus, where the SH0ES+Pantheon+ constraint on H$_0$ cannot grow arbitrarily large, even if the Pantheon+ inter-survey zeropoints are grossly miscalibrated, the accuracy of the CCHP constraint on H$_0$ is entirely dependent on the accuracy of the photometric calibration between CSP and the remaining TRGB calibrator surveys. Taking the conservative but not outlandish estimation of $\sigma_{P,S} \simeq 25$ mmag used throughout this work, the additional H$_0$ uncertainty that CCHP might experience as a result of inter-survey photometric miscalibration photometric could be as large as $0.77$ km s$^{-1}$ Mpc$^{-1}$.  

\subsection{Conclusion}

In this paper, we assessed the extent to which miscalibrated zeropoint offsets between the surveys of the Pantheon+ SN Ia compendium, $\Delta \mu_S$, could degrade constraints on measurements of cosmological parameters.   
We determined that (effectively) freely floating photometric zeropoint offsets, $\Delta \mu_S$, could contribute an additional statistical uncertainty of $0.15$ km s$^{-1}$ Mpc$^{-1}$ to SH0ES+Pantheon+ measurements of H$_0$, but could increase the SH0ES+Pantheon+ joint uncertainty on $\Omega_M$ and $w$ by as much as 150\%.
Those floating $\Delta \mu_S$ could shift the best-fit values of H$_0$, $\Omega_M$, and $w$ by $0.35$km s$^{-1}$ Mpc$^{-1}$, 0.12, and 0.5, respectively.  
Notably, we determined that the systematic uncertainty contributions of the $\Delta \mu_S$ could grow no larger than these maximum values, as larger inter-survey zeropoint offset uncertainties do not improve the correspondence between the cosmological fit and the Pantheon+ data.  
We explained the reasons for this plateauing in Appendix \ref{app:H0SelfCalibration}. 

Focusing specifically on the Hubble Tension (HT) between SH0ES SNe Ia distance ladder measurements of H$_0$ and Planck $\Lambda$CDM-based inferences H$_0$, we found that uncertainties in the inter-survey zeropoint offsets could reduce this tension by only $0.03\sigma$.  
Gray miscalibrations between Pantheon+ surveys cannot account for HT, at least not alone.

Our findings suggest that SNe Ia compendium-based distance ladder measurements of cosmological parameters benefit from an effective self-calibration of gray systematics between surveys, capping the inter-survey systematic error in compendium-based measurements of cosmological parameters.  
However, this self-calibration occurs only when there are numerous surveys that span both the middle and Hubble Flow (HF) rungs of the cosmic distance ladder.  
Programs that leverage only one HF survey, such as CCHP, are more susceptible to undetected inter-survey systematic than those that include many HF surveys.  

We focused exclusively on gray inter-survey calibration errors.  Although chromatic cross-survey photometric calibration errors, owing to distinct filter systems, chromatic telescope throughputs, etc., are a matter of legitimate concern, the SNe Ia processing pipeline does not presently lend itself to varying the chromatic inter-survey photometry.  Such chromatic differences are embedded in the SNe Ia lightcurve fitting algorithms, making the prospect of varying them significantly more complicated than the fixed, post-lightcurve fitting $\mu$ offsets that we considered here.  
A future, more in-depth analysis could take up the task of measuring the effects of chromatic inter-survey calibration errors by varying the SNe light curve parameters as part of the larger MCMC sampling loop.  

Accurate inter-survey calibrations in compendia of SNe Ia are still extremely important.  
SH0ES+Pantheon+ measurements of $\Omega_M$ and $w$ are reliable only if the inter-survey zeropoints are well-calibrated.  
Further, the additional cosmic constraining power achieved when additional surveys are added to the Pantheon+ compendium is comparable to the greater accuracy achieved by improving the inter-survey zeropoint calibration (see Section \ref{sec:relStats}).     
As more surveys of SNe Ia are gathered, effort should be taken to integrate them into the Pantheon+ (or a future larger) compendium, and the relative survey flux scales should be calibrated with utmost accuracy.  

The creation of a well-calibrated SNe Ia compendium is a monumental undertaking, and our results underscore how valuable such compendia are in measuring the recent expansion history of the Universe.  
Even if a SNe compendium suffers from undiagnosed inter-survey calibration errors, it can still provide meaningful cosmic constraints, especially on H$_0$.   
We particularly demonstrate that the SH0ES+Pantheon+ measurement of H$_0$ is degraded by $<1$\% even for unrealisticaly larger inter-survey calibration errors, and thus that cross-survey calibration uncertainty does not represent a substantial impediment to the longstanding objective of measuring H$_0$ to within 1\%.  
The HT between Cepheid + SNe Ia distance ladder measurements of H$_0$ and Planck $\Lambda$CDM inferences of H$_0$ is not ascribable to gray inter-survey miscalibration, and is not likely to diminish with additional surveys of even hundreds of spectroscopically matched SNe Ia.  

However, we do find that the SH0ES+Pantheon+ measurements of $\Omega_M$ and $w$ are highly sensitive to the accuracy of inter-survey cross-calibration.  Our results emphatically do not demonstrate that the SH0ES+Pantheon+ cosmology is inaccurate, but rather underscores the immense importance of the painstaking cross-survey calibration performed by the Pantheon+ analysis team.  
Only a well orchestrated, all sky campaign to observe thousands of spectroscopically-matched SNe on a single photometric system, such as the ambitious SNe Ia observing program of the Vera Rubin Observatory,  \citep{Ivezic2019}, could use SNe Ia to measure $\Omega_M$ and $w$ free of possible cross-survey calibration errors.  

\section{Acknowledgements} 
We are grateful to the US Department of Energy for their support of our assessments of the cosmic constraints offered by SNe Ia, under DOE grant DE-SC0007881, and to Harvard University for its support of our program.
Dillon Brout acknowledges support for this work was provided by NASA through the NASA Hubble Fellowship grant HST-HF2-51430.001 awarded by the Space Telescope Science Institute, which is operated by Association of Universities for Research in Astronomy, Inc., for NASA, under contract NAS5-26555.
This work was completed in part with resources provided by the Harvard University FAS Research Computing and University of Chicago Midway Research Computing Center.
SN distance determination pipeline managed by \texttt{PIPPIN} \citep{Pippin}. 
D.S. is supported by DOE grant DE-SC0010007 and the David and Lucile Packard Foundation. D.S. is supported in part by the National Aeronautics and Space Administration (NASA) under Contract No. NNG17PX03C issued through the Roman Science Investigation Teams Programme.


\appendix 

\section{Effects of Inter-Survey Zeropoint Offsets on SNe Ia Distance Moduli } \label{app:DeltaMuDerivation}  

Fundamentally, adjusting distance moduli represents an adjustment of the supernova flux from an initial value, $f_{init}$, to a shifted value, $f_{shift}$: 
\beq \label{eq:shiftedFlux}
f_{shift} = f_{init} 10 ^ {- \Delta \mu_S / 2.5} \ .
\eeq 
By adjusting the distance moduli in Survey $S$, we conjecture that the fluxes of the SNe in that survey are greater than initially reported (if $\Delta \mu_S > 0$) or less than initially reported (if $\Delta \mu_S < 0$).  

Adjusting the fluxes of Cepheid-paired SNe also changes our measurement of the normalized SN Ia absolute luminosity from an initial value, $L_{init}$, to a shifted value, $L_{shift}$.  As we will see, the logarithmic ratio of this adjusted luminosity introduces an additional distance modulus offset to the distance moduli of the Hubble Flow SNe. 
For Cepheid-paired supernova $i$, this luminosity-sourced distance modulus offset is: 
\beq
 \Delta \mu_{L,i} = 2.5 \log_{10}{\frac{L_{shift, i}}{L_{init, i}}}  = 2.5 \log_{10}{\frac{L_{ceph} f_{shift, i}/f_{ceph} }{L_{ceph} f_{init, i}/f_{ceph} } }  = 2.5 \log_{10}{\frac{f_{shift, i} }{f_{init, i}} }  = 2.5 \log_{10} (10 ^ {-\Delta \mu_S/2.5}) = -\Delta \mu_S \ , 
\eeq 
where we used the fact that the ratio of the initial and shifted fluxes is the exponential of the appropriate distance modulus residual survey offset (Equation \ref{eq:shiftedFlux}) and where $L_{ceph}$ and $f_{ceph}$ are respectively the absolute luminosity and measured flux of the Cepheid star to which SNe $i$ is tied. 
The best-fit inferred value of this luminosity distance modulus offset, $\Delta \mu_L$, is the weighted mean of all the individual luminosity distance modulus offsets: 
\beq \label{eq:fullDeltaMu_app}
\Delta \mu_L =  \frac{\sum_{i=1}^{N_{Ceph}} \Delta \mu_{L,i} \sigma_{\mu,i}^{-2} } {\sum_{i=1}^{N_{Ceph}} \sigma_{\mu,i}^{-2} } = - \frac{\sum_{i=1}^{N_{Ceph}} \Delta \mu_{S,i} \sigma_{\mu,i}^{-2} } {\sum_{i=1}^{N_{Ceph}} \sigma_{\mu,i}^{-2} } \ ,
\eeq
where $\sigma_{\mu, i}$ is the reported uncertainty in the distance modulus of SN $i$, $\Delta \mu_{S,i}$ is the distance modulus offset applied to the survey containing SN $i$, and the sum is taken only over Cepheid-paired SNe.  


The initially measured distance modulus for Hubble Flow supernova $i$, $\mu_{init,i}$, is derived from $f_{init,i}$ and $L_{init,i}$: 
\beq
\mu_{\textrm{meas, init}, i} = 25 + 5 \log_{10}{\Big ( \sqrt{\frac{L_{\mathrm{init}} }{4 \pi f_{\mathrm{init},i} }} \mathrm{Mpc}^{-1} \Big ) }  \ . 
\eeq 
Applying inter-survey distance modulus offsets, $\Delta \mu_S$, to the entire reported set of Pantheon+ distance moduli (both Cepheid and Hubble Flow) adjusts the measured Hubble Flow distance moduli in two ways: by changing the measured flux and by changing the measured value of the SN Ia luminosity.  
The shifted distance modulus of Hubble Flow supernova $i$, $\mu_{\textrm{meas, shift, }i}$, in survey $S$ is thus: 
\beq
& \mu_{\textrm{meas, shift, }i} = 25 + 5 \log_{10}{\Big ( \sqrt{\frac{L_{\mathrm{shift}} }{4 \pi f_{\mathrm{shift},i} }} \mathrm{Mpc}^{-1} \Big ) }   \\ 
& = 25 + 5 \log_{10}{\Big ( \sqrt{\frac{L_{\mathrm{init}} }{4 \pi f_{\mathrm{init},i} }} \mathrm{Mpc}^{-1} \Big ) }  \ \ \ + 2.5 \log_{10}{\Big ( \frac{L_{\mathrm{shift}} }{L_{\mathrm{init} }}  \Big ) } + 2.5 \log_{10}{\Big ( \frac{f_{\mathrm{init},i} }{f_{\mathrm{shift},i }} \Big ) }  = \mu_{\textrm{meas, init}} + \Delta \mu_L + \Delta \mu_S
\ . 
\eeq 
As we discuss further in Section \ref{app:H0SelfCalibration}, because $\Delta \mu_L$ is the average inter-survey zeropoint offset of all Cepheid paired SNe, $\Delta \mu_L + \Delta \mu_S$ only shifts the zeropoint of survey $S$ relative to the average shift of surveys with Cepheid-paired SNe.  

\section{Self-Correcting Effect of Cepheid-paired SNe Data Sets in Determining Cosmological Parameters } \label{app:H0SelfCalibration}

\begin{figure*}
\centering
\includegraphics[width=1.0\textwidth]{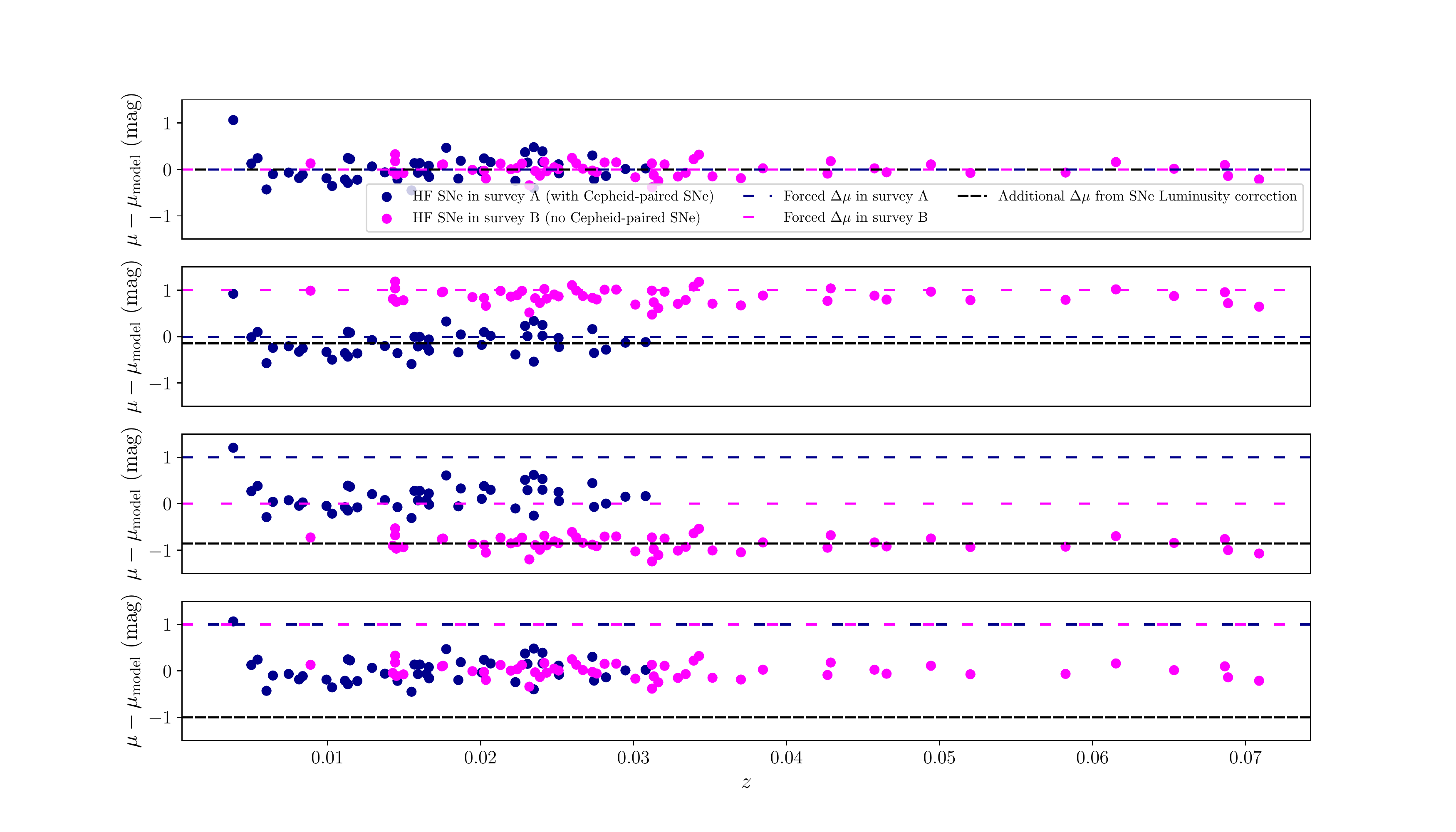}
\caption{\label{fig:SurveyASurveyB} A simple artificial SNe Ia data set with two surveys: one of which (Survey A, blue) includes Cepheid paired and Hubble flow SNe Ia, while the other (Survey B, pink) has only Hubble flow SNe Ia.  From top to bottom, we introduce the following inter-survey zeropoint offsets to these data: no offset, an offset of 1 magnitude to survey B, an offset of 1 magnitude to survey A, and a 1 magnitude offset to both surveys.  The values of the offsets for Surveys A and B are indicated with dashed lines of the appropriate color.  Because only Survey A includes Cepheid-paired SNe, the luminosity offset, $\Delta \mu_L$ indicated with a black dashed line, is exactly opposite the applied distance modulus offset of Survey A (see Equation \ref{eq:fullDeltaMu}).  According to Equation \ref{eq:DeltaMu}, all surveys are shifted by $\Delta \mu_L$.  Thus, applying a zeropoint offset to Survey A does not shift Survey A, but effectively introduces the opposite shift to Survey B.  Inter-survey offsets can only separate Surveys A and B, but cannot shift the surveys collectively.  
Therefore, the relative survey offsets can grow only so large before they become incompatible with a single cosmic fit.  This effect is why, in Figure \ref{fig:H0PostOnGaussPrior}, the posterior uncertainties of H$_0$, $\Omega_M$, and $w$ plateau at some threshold $\Delta \mu_S$ prior width.  }
\end{figure*}

In Figure \ref{fig:PostSigOnGaussPrior}, the widths of the H$_0$ and $\Omega_M \times w$ posterior distributions do not increase with $\Delta \mu_S$ when $\Delta \mu_S > 200$ mmag.   
This occurs because, as we show in Equation \ref{eq:DeltaMu}, the updated SN Ia luminosity introduces an additional distance modulus offset, $\Delta \mu_L$.  This luminosity offset is opposite in sign and equal in magnitude to the weighted average of the inter-survey zeropoints of all Cepheid paired SNe (Equation \ref{eq:fullDeltaMu}).  
Therefore, any $\Delta \mu_S$ offset shared between surveys has no effect on the $\chi^2$ statistic because $\Delta \mu_L$ exactly cancels it.  Differential $\Delta \mu_S$ offsets are not cancelled and pull the SNe distance moduli vertically apart in the Hubble Diagram.  As these differential offsets become larger, the distance moduli residuals become less compatible with a single set of cosmological parameters. 

To illustrate this phenomenon, we consider a simplified SNe compendium that consists of only two surveys: Survey A, which contains both Hubble Flow and Cepheid paired SNe Ia, and survey B which contains only Hubble Flow SNe Ia.  
As we show in Figure \ref{fig:SurveyASurveyB}, applying an inter-survey offset to only one survey pulls the $\Delta \mu$ values of each survey vertically apart, while applying the same offset to both surveys produces no net change in $\Delta \mu$.  
Thus, the best-fit value of H$_0$ cannot change without limit, even for arbitrarily large inter-survey offsets.
   
This cancellation of $\Delta \mu_L$ with the average inter-survey zeropoint offsets of Cepheid-paired SNe is perfect only because we assume (incorrectly) that the distance moduli of the Cepheid-calibrators are known exactly.  A more complete analysis of this data would include additional degrees of freedom to reflect the uncertainty of HST measurements of Cepheid fluxes.  Adjustments to the Cepheid distance moduli would add an additional term to Equation \ref{eq:fullDeltaMu}, partially decoupling $\Delta \mu_L$ from the average $\Delta \mu_S$ of Cepheid-paired SNe.


\bibliography{VariableMuOffsetBibliography}{}
\bibliographystyle{aasjournal}



\end{document}